\documentclass[letterpaper,aps,pre,twocolumn,showpacs,preprintnumbers,
superscriptaddress]{revtex4-1}

\usepackage{graphicx}
\usepackage{epsf} 
\usepackage{epsfig}
\usepackage{dcolumn}
\usepackage{bm}
\usepackage{amssymb}
\usepackage{amsmath}
\usepackage{amsbsy}

\usepackage{graphicx,color,ifthen}

\begin{document}

\title{Kardar-Parisi-Zhang asymptotics for the two-dimensional noisy Kuramoto-Sivashinsky equation}

\author{Matteo Nicoli}
\affiliation{Laboratoire de Physique de la Mati\`ere Condens\'ee, \'Ecole Polytechnique - CNRS,
91128 Palaiseau, France}

\author{Edoardo Vivo}
\author{Rodolfo Cuerno}
\affiliation{Departamento de Matem\'aticas and Grupo Interdisciplinar de
 Sistemas Complejos (GISC), Universidad Carlos III de Madrid, Avenida de la
 Universidad 30, E-28911 Legan\'es, Spain}

\date{\today}

\begin{abstract}
We study numerically the Kuramoto-Sivashinsky (KS) equation forced by external white noise in two space dimensions, that is a generic model for e.g.\ surface kinetic roughening in the presence of morphological instabilities. Large scale simulations using a pseudospectral numerical scheme allow us to retrieve Kardar-Parisi-Zhang (KPZ) scaling as the asymptotic state of the system, as in the 1D case. However, this is only the case for sufficiently large values of the coupling and/or system size, so that previous conclusions on non-KPZ asymptotics are demonstrated as finite size effects. Crossover effects are comparatively stronger for the 2D case than for the 1D system.
\end{abstract}


\mbox{Published as Phys.\ Rev.\ E {\bf 82}, 045202(R) (2010).}

\maketitle

The Kuramoto-Sivashinsky (KS) equation is a paradigmatic model for chaotic spatially extended systems, arising in a variety of physical contexts, like thin solid films, interfaces between viscous fluids, waves in plasmas and chemical reactions, reaction-diffusion systems, or combustion fronts \cite{cross:2009}. Actually, in its stabilized form, it has been shown to provide a generic model for parity-symmetric systems featuring a bifurcation with a vanishing wave number \cite{misbah:1994}. In the presence of external fluctuations, a natural generalization is provided by the {\em noisy} KS (nKS) equation, that reads
\begin{equation}
\frac{\partial h}{\partial t} = - \nu \nabla^2 h - \mathcal{K} \nabla^4 h + \frac{\lambda}{2} (\nabla h)^2 + \eta(\mathbf{r},t),
\label{nks}
\end{equation}
where $\mathbf{r}\in\mathbb{R}^d$, we will take $\nu$, $\mathcal{K}$, $\lambda$ to be {\em positive} parameters, and $\eta(\mathbf{r},t)$ is a Gaussian white noise with zero mean and correlations
\begin{equation}
\langle \eta(\mathbf{r},t) \eta(\mathbf{r}',t') \rangle = 2 D \delta(\mathbf{r}-\mathbf{r}') \delta(t-t') .
\label{eta}
\end{equation}
Thus, the nKS equation reduces to the deterministic KS (dKS) equation in the $D = 0$ case. Indeed, Eqs.\ \eqref{nks}-\eqref{eta} appear in a wide variety of physical contexts, from e.g.\ step dynamics in epitaxy \cite{karma:1993} to surface erosion by ion-beam sputtering (IBS) \cite{cuerno:1995a}, or diffusion-limited growth \cite{cuerno:2001}. In these, $h(\mathbf{r},t)$ can be thought of as the position at time $t$ of a moving front above point $\mathbf{r}$ on a reference line or plane, that will be the physical image to be used in this work. One of the intriguing features of the dKS equation is the fact that, at least for $d=1$ \cite{krug:1997}, its large scale properties display kinetic roughening in the universality class of the (stochastic) Kardar-Parisi-Zhang (KPZ) equation, as also occurs for the nKS equation \cite{cuerno:1995b,cuerno:1995c}. This links the two seemingly opposite phenomena of pattern formation and scale invariance within the evolution of a single system at appropriate time and length scales. However, the two-dimensional case $d=2$ remains controversial: on the one hand, there are opposing claims \cite{Procaccia:1992,Jayaprakash:1993} on the asymptotics of the dKS equation vs that of the 2D KPZ equation; on the other hand, the asymptotics of the 2D nKS equation is not well understood. In $d=1$ the nKS equation indeed belongs to the KPZ universality class, as borne out from numerical simulations \cite{cuerno:1995b} and dynamic renormalization group (DRG) analysis \cite{cuerno:1995c}. However, as for the KPZ equation, the DRG approach is inconclusive in $d=2$. Numerical results \cite{Drotar:1999} suggest non-KPZ asymptotics, contradicting na\"{\i}ve expectations based on the structure of the RG flow, in which $\nu$ seems to change sign under renormalization as in $d=1$ \cite{cuerno:1995c}.

In this work, we revisit the numerical study of the 2D nKS equation. Using an improved numerical scheme, we perform large scale simulations that allow us to identify KPZ scaling as the asymptotic state. However, this only occurs for sufficiently large values of the effective coupling in the system and/or system size, previous conclusions on non-KPZ asymptotics being due to finite size effects. Nevertheless, crossovers are comparatively stronger for the 2D case than for the 1D system, which has possibly prevented earlier works from assessing the actual hydrodynamic behavior. Our results may guide in the assessment of the large scale behavior of physical systems described by the 2D nKS equation.

Note that the nKS system \eqref{nks}-\eqref{eta} depends on a single free parameter; for instance, by rescaling $\mathbf{r}\to (\mathcal{K}/\nu)^{1/2}\, \mathbf{r}$, $t\to(\mathcal{K}/\nu^2)\, t$, and $h(\mathbf{r},t) \to(D/\nu)^{1/2} \, h(\mathbf{r},t)$, it can be written as
\begin{equation}
\frac{\partial h}{\partial t} = - \nabla^2 h - \nabla^4 h + \dfrac{\sqrt{g}}{2} (\nabla h)^2 + \xi(\mathbf{r},t) ,
\label{nks1}
\end{equation}
where $g = \lambda^2 D / \nu^3$ and the rescaled noise $\xi(\mathbf{r},t)$ has zero mean and variance
\begin{equation}
\langle \xi(\mathbf{r},t) \xi(\mathbf{r}',t') \rangle =  2 \, \delta(\mathbf{r}-\mathbf{r}') \delta(t-t') .
\end{equation}
Thus, we can study the full phase space of the nKS equation as a function of only the coupling constant $g$ and the lateral system size $L$.

Initially motivated by results for IBS, Drotar {\em et al.} \cite{Drotar:1999} solved numerically \eqref{nks}-\eqref{eta} for several values of the parameters $\nu$, $\mathcal{K}$ and $\lambda$. They pointed out the importance of the coupling $g$, but were unable to reach a well defined asymptotic regime. In fact, they found two different scaling regimes, in terms of the exponents values determined from the behavior of the surface roughness $W^2(t) = \langle (1/L^2)\sum_{\mathbf{r}}[h(\mathbf{r},t)-\bar h]^2\rangle$ and height-difference correlation function $G(\mathbf{r},t)=\langle (h(\mathbf{r}^{\prime}+\mathbf{r},t)-h(\mathbf{r}^{\prime},t))^2\rangle$; here, bar denotes space average and brackets denote ensemble average. Thus, in the presence of kinetic roughening \cite{Barabasi:1995}, the roughness scales as $W \sim t^{\beta}$ before reaching the stationary state, after which $W \sim L^{\alpha}$, where $\beta$ and $\alpha$ are called growth and (global) roughness exponents, respectively. Moreover, $G(r,t) \sim t^{2\beta}$ [$G(r,t) \sim r^{2\alpha_{loc}}$] for $r\gg t^{1/z}$ [$r\ll t^{1/z}$], where $z=\alpha/\beta$ is the dynamic exponent, and $\alpha_{loc}=\alpha$ for the standard Family-Vicsek behavior, while $\alpha_{loc}\neq\alpha$ in the presence of so-called anomalous scaling \cite{krug:1997}.

Taking into account the uncertainties of the estimates in \cite{Drotar:1999} and spurious effects due to the crossover between the two scaling regimes found, the results obtained in this reference for the early stage of the growth process are compatible with the Mullins-Herring fixed point \cite{Barabasi:1995}. This is consistent with the anomalous scaling found for the height-height correlation function (see Fig.\ 11 in \cite{Drotar:1999} for $t < 50$). For late times, Drotar {\em et al.} showed that surfaces produced by the nKS equation display scale invariance with exponent values for $\alpha$ and $\beta$ in the ranges $0.25-0.28$ and $0.16-0.21$, respectively. These values differ substantially from those associated with the KPZ universality class for $d=2$, even allowing for the spread that the latter have \cite{canet:2010}. For the sake of homogeneity, we will take as reference values for the exponents those we obtain for the 2D KPZ equation with the same numerical procedure that will be subsequently employed for the nKS equation (see below). These are \cite{nicoli:unpubl} $\alpha_{\rm KPZ} = 0.39 \pm 0.01$ and $\beta_{\rm KPZ}=0.24\pm 0.01$.

An important remark concerns the numerical scheme used for the integration of \eqref{nks}-\eqref{eta}. Drotar {\em et al.} chose a standard finite-difference discretization for space derivatives. Currently it is accepted that such a scheme underestimates both the KPZ nonlinearity and the effective coupling $g$ \cite{Gallego:2007,Wio:2010_all}. Here we opt for a pseudo-spectral scheme that has been successfully used for the numerical integration of local \cite{Gallego:2007,Giada:2002} and non-local \cite{Nicoli:2008_all} stochastic equations featuring nonlinearities of the KPZ type. Details of this numerical method can be found e.g. in \cite{Canuto:1987,Giacometti:2000,Gallego:2007}.

We start by considering parameter values of the nKS equation \eqref{nks}-\eqref{eta} that correspond to relatively small coupling values from $g=2 \cdot 10^{-2}$ up to $g=2 \cdot 10^{3}$. We achieve this by tuning $\lambda$ while keeping other parameters fixed at $L=512$,  $\nu = 0.2$, $\mathcal{K} = 2$, $D = 1$, $\Delta x = 2$, and $\Delta t=5\cdot 10^{-3}$, the latter being the lattice spacing and the time step, respectively. For each parameter set, we measure the global surface roughness $W(t)$ and the (circular average of) the power spectral density (PSD) or height structure factor \cite{Barabasi:1995} $S(\mathbf{k}) = \langle\hat h_{\mathbf k}\hat h_{-\mathbf{k}}\rangle$ as functions of time. Here, $\hat h_{\mathbf k}(t)$ is the 2D Fourier transform of $h(x,t)-\overline{h}(t)$. In these simulations, observables are averaged over $50$ noise realizations. The structure factor has been shown to feature more clear scaling behavior than real-space correlations in the presence of crossover effects \cite{siegert:1996}, that are expected here.
\begin{figure}[t!]
\begin{center}
\epsfig{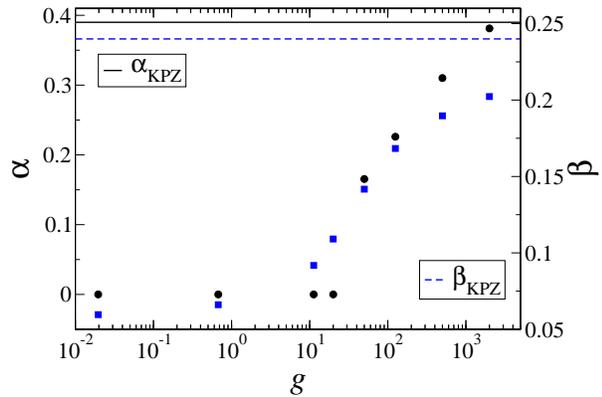}
\caption{(Color online) Exponent values of the 2D nKS equation as functions of $g$, in the weak coupling regime. Solid black bullets (ordinates on left vertical axis) provide values of the roughness exponent $\alpha$; blue squares (ordinates on right vertical axis) provide values of the growth exponent $\beta$. The solid black and dashed blue lines indicate our reference values for the exponents of the 2D KPZ equation, $\alpha_{\rm KPZ} = 0.39$ and $\beta_{\rm KPZ}=0.24$, respectively.
}
\label{fig:trans_exp}
\end{center}
\end{figure}

By fitting the long time behavior of $W(t)$ prior to saturation and the small $k=|\mathbf{k}|$ behavior $S(k) \sim 1/k^{2\alpha+2}$ at the stationary state \cite{Barabasi:1995}, we find $\beta$ close to 0 (log) for $g\lesssim 10$, increasing up to $\beta=0.20\pm 0.01$ for $g= 2 \cdot 10^{3}$, see Fig.\ \ref{fig:trans_exp}. The roughness exponent $\alpha$ is also consistent with 0 (log) for $g\lesssim 20$, after which it increases, reaching up to $\alpha = 0.39\pm 0.01$ for $g= 2000$, see Fig.\ \ref{fig:trans_exp}. Thus we conclude that, even at small couplings, the KPZ nonlinearity is able to tame the linear instability in the nKS equation and induce kinetic roughening properties. For $g\lesssim 10$ these are in the (2D) Edwards-Wilkinson (EW) universality class \cite{Barabasi:1995}, as for the 1D nKS case \cite{cuerno:1995b,cuerno:1995c}. However, for larger coupling values the scaling behavior is neither EW nor KPZ, although the value of $\alpha$ for $g=2000$ seems already reminiscent of KPZ behavior.
Since EW and KPZ scaling are precisely the two meaningful fixed points that are found through DRG analysis \cite{cuerno:1995c}, we believe that the intermediate exponent values found in Fig.\ \ref{fig:trans_exp} are to be thought of as non-asymptotic behavior due to the finite system size of our simulations. This behavior is analogous to that of the 1D nKS equation \cite{cuerno:1995b,cuerno:1995c}.

\begin{figure}[t!]
\begin{center}
\epsfig{file=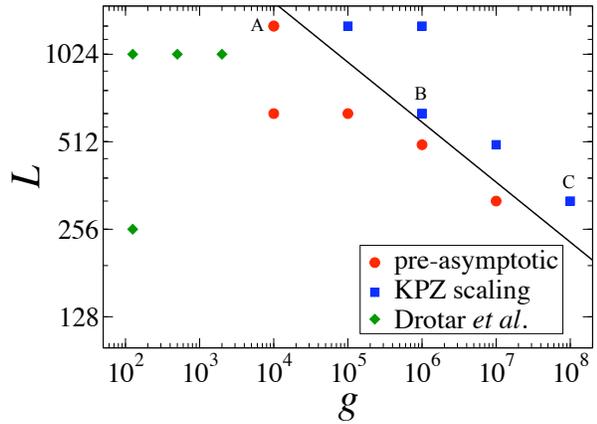,width=0.475\textwidth,clip=}
\caption{(Color online) Qualitative asymptotic scaling of the nKS equation (see main text) in $(g,L)$ parameter space. In simulations, $\nu = 0.1$ and $\mathcal{K}=4$, $\Delta x=1.25$, and $\Delta t=5\cdot 10^{-3}$ are fixed, different values of $g$ being achieved by tuning $\lambda$ and $D$. Red bullets correspond to pre-asymptotic scaling (type i behavior) while blue squares correspond to strong coupling, KPZ scaling (type ii and iii behaviors). As examples, points A, B, and C are explicitly discussed in \cite{epaps}. Green diamonds are results from \cite{Drotar:1999}. The solid line is a reference fit computed by least squares, separating preasymptotic from asymptotic scaling behavior.}
\label{fig:kpz-uc}
\end{center}
\end{figure}

In order to confirm this interpretation, we need to explore larger coupling and/or system size values. By increasing $g$, indeed we have been able to reach an asymptotic state in which the critical exponents are compatible with those of the 2D KPZ universality class.
\begin{figure}[t!]
\begin{center}
\epsfig{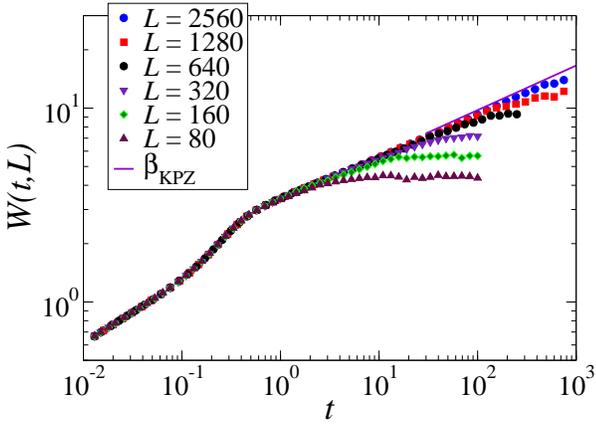}
\caption{(Color online) Time evolution of the surface roughness for different values of the system size $L$.
For all simulations we use $\nu=0.1$, $\mathcal{K}=4$, $\lambda=20$, $D=50$, $\Delta x =1.25$, and $\Delta t =5\cdot10^{-3}$,
leading to $g=2\cdot 10^7$. Results have been averaged over a different number of noise realizations:
$15$ for $L=2560$, $30$ for $L=1280$, and $100$ for the remaining values of $L$ in the
legend. The purple solid line is a guide for the eye, and has slope $\beta_{\rm KPZ}=0.24$.
}
\label{fig:w-strong}
\end{center}
\end{figure}
As seen above, the roughness exponent $\alpha$ already reaches a KPZ-compatible value already for {\em moderate} values of $g$ and $L$. However, $\beta$ approaches its asymptotic KPZ value only very slowly. The fact that $\alpha$ reaches its asymptotic value earlier (i.e., for smaller $g$ values) than $\beta$ has been also reported in other studies of crossover phenomena within kinetic roughening \cite{siegert:1996}, and may be partially accounted for by the exact link between the roughness and the PSD, $W^2(L,t) = \sum_{\mathbf{k}} S(\mathbf{k},t)$. Unambiguous assessment of the asymptotic $\beta$ value is only possible for {\em large} $g$ and $L$, as seen in Fig.\ \ref{fig:w-strong}, in which $W(t)$ is plotted for several system sizes at a fixed large coupling value $g=2\cdot 10^7$. This very long crossover between pre-asymptotic and asymptotic states hinders the possibility to reach the strong coupling KPZ fixed point for small system sizes and small values of the coupling, which applies to the simulations previously reported for this system \cite{Drotar:1999}, see green diamonds in Fig.\ \ref{fig:kpz-uc}.
In order to analyze the situation in more detail, we have estimated $\beta$ with a more robust methodology. For kinetic roughening systems, we can obtain the roughness exponent and the dynamic exponent simultaneously from a standard data collapse of the power spectral density \cite{Barabasi:1995}, in which we rescale $k\to k \, t^{1/z}$ and $S \to S \, k^{2\alpha+2}$.  Therefore, an investigation of the strong coupling regime has been carried out for representative points in the $(g,L)$ plane by collapsing the corresponding PSD functions using 2D KPZ exponents, see \cite{epaps} for some specific examples. Note, in our case we are interested in asymptotic scaling so that collapse with such exponents values is only to be expected for the smallest $k$'s in the system. Following this procedure, we have been able to identify three different behaviors for a fixed $g$ and increasing substrate size $L$: i) Pre-asymptotic regime in which the collapse of the PSD is poor for every value of $L$ in our range (see Figs.\ 1a-1b in \cite{epaps}, point A in Fig.\ \ref{fig:kpz-uc}); ii) regime in which only the lowest wave-numbers (very large wavelengths) of the PSD collapse with the KPZ exponents (Figs.\ 2a-2b in \cite{epaps}, point B in Fig.\ \ref{fig:kpz-uc}); and iii) fully developed strong coupling behavior in which KPZ asymptotics is reached immediately after the exponential growth (due to the linear instability) of the surface roughness (Figs.\ 3a-3b in \cite{epaps}, point C in Fig.\ \ref{fig:kpz-uc}). Results are qualitatively summarized in Fig.\ \ref{fig:kpz-uc}. In this figure we can see that, for small $g<10^4$ and $L\simeq 1024$, we have not yet been able to reach the KPZ regime (type i behavior). Actually, in the $g\to 0$ limit the nKS equation becomes linear (and ill-posed) and KPZ behavior does not occur for any $L$ value. We then find an intermediate region (for $10^5 \leq g \leq 10^7$ and $512\leq L \lesssim 1024$) in which dynamic crossover behavior occurs between the pre-asymptotic regime and KPZ scaling, albeit the latter only applies to the largest accesible scales (type ii behavior). Finally, for $g > 10^7$, 2D KPZ scaling is readily observed even for small systems, $L\simeq 256$ (type iii behavior).

In summary, our numerical results show that the nKS equation is asymptotically in the KPZ universality class in two space dimensions, confirming previous expectations derived from RG analysis, and generalizing known results in $d=1$ to one higher dimension. This result moreover can guide the interpretation of large scale experimental and/or numerical data obtained in the different contexts for which this equation appears as a physical model. Note that, even in experimental systems, crossover effects may hinder observation of actual asymptotic behavior at accesible scales, see \cite{Nicoli:2008_all} and references therein. Moreover, and also of practical implications, crossover effects are substantially stronger for the 2D case than for the 1D case, in the sense that, fixing all parameter values including the system size $L$, the nKS equation can be already in the KPZ asymptotic state for $d=1$ while only preasymptotic scaling can be measured for $d=2$. As an example, Ueno {\em et al.} \cite{cuerno:1995c} obtain KPZ scaling for the 1D nKS equation already at $g=20$ (and $L=2\cdot 10^4$), while in our 2D case this value of $g$ leads to non-asymptotic scaling for any feasible system size, see Fig.\ \ref{fig:trans_exp}. This 1D vs 2D difference might be due to the particularly strong effect that fluctuations have in one dimension, which may aid the approach to the stationary state for a given parameter set.

In the context of the controversy on the universality class of the {\em deterministic} 2D KS (dKS) equation, if an effective description of the dKS equation by an ``equivalent'' nKS equation were achieved as in the 1D case \cite{chow:1995}, then our results would imply that the asymptotic scaling of the 2D dKS equation is in the 2D KPZ class. However, such a link is not yet available for the $d=2$ case \cite{Boghosian:1999}, and in the absence of further progress in that direction the controversy remains an important open quesion in Nonlinear Science. Reflecting on the complexity of this problem, one may draw lessons from the case of the related
Michelson-Sivashinsky (MS) equation, that is a model for e.g.\ flame front propagation \cite{bychkov:2000}. Actually, both the KS and the MS equations take a very similar shape in $\mathbf{k}$ space, thus
\begin{equation}
\partial_t h_{\mathbf{k}} = (\nu k^n - {\cal K} k^m) h_{\mathbf{k}} + \frac{\lambda}{2} {\cal F}[(\nabla h)^2] ,
\label{nm_eq}
\end{equation}
where ${\cal F}[\cdot]$ stands for space Fourier transform, $(n,m)=(2,4)$ for the KS equation and $(n,m)=(1,2)$ for the MS equation \cite{Nicoli:2008_all}. The asymptotic states of the deterministic MS equation and of the noisy MS equation [obtained by adding a noise term to the rhs of Eq.\ \eqref{nm_eq}, much like the nKS generalizes the dKS equation] are known to be quite different \cite{karlin:2004}. Nevertheless, this difference turns out to be hard to assess in practice, as unavoidable numerical noise (round-off errors) in any numerical integration of the deterministic MS equation has been seen to transform the problem into that of its stochastic generalization \cite{karlin:2004}. A similar ``practical" difficulty in telling properties of the deterministic equation apart from those of the stochastic generalization may apply in the 2D KS context, although whether that is the case remains to be seen in the future.

Partial support for this work has been provided by MICINN (Spain) Grant No.\ FIS2009-12964-C05-01. E.\ V.\ acknowledges support by Universidad Carlos III de Madrid through a predoctoral scolarship.


\begin{thebibliography}{99}

\bibitem{cross:2009} M. Cross and H. Greenside, {\em Pattern Formation and Dynamics in Nonequilibrium Systems} (Cambridge University Press, Cambridge, England, 2009).

\bibitem{misbah:1994} C. Misbah and A. Valance, Phys. Rev. E {\bf 49}, 166 (1994).

\bibitem{karma:1993} A. Karma and C. Misbah, Phys. Rev. Lett. {\bf 71}, 3810 (1993).

\bibitem{cuerno:1995a} R. Cuerno and A.-L. Barab\'asi, Phys. Rev. Lett. {\bf 74}, 4746 (1995).

\bibitem{cuerno:2001} R. Cuerno and M. Castro, Phys. Rev. Lett. {\bf 87}, 236103 (2001).

\bibitem{krug:1997} J. Krug, Adv. Phys. {\bf 46}, 139 (1997).

\bibitem{cuerno:1995b} R. Cuerno, H. A. Makse, S. Tomassone, S. T. Harrington, and H. E. Stanley, Phys. Rev. Lett. {\bf 75}, 4464 (1995).

\bibitem{cuerno:1995c} R. Cuerno and K. B. Lauritsen, Phys. Rev. E {\bf 52}, 4853 (1995); K. Ueno, H. Sakaguchi, and M. Okamura, {\em ibid.} {\bf 71}, 046138 (2005).

\bibitem{Jayaprakash:1993}
C. Jayaprakash,  F. Hayot,   and R. Pandit, Phys. Rev. Lett. {\bf 71}, 12 (1993).

\bibitem{Procaccia:1992}
I. Procaccia, M. H. Jensen, V. S. L'vov, K. Sneppen,  and R. Zeitak, Phys. Rev. A {\bf 46}, 3220 (1992).

\bibitem{Drotar:1999}
J. T. Drotar, Y.-P. Zhao, T.-M. Lu, and G.-C. Wang, Phys. Rev. E {\bf 59}, 177 (1999).

\bibitem{Barabasi:1995}
A.-L. Barab\'asi and H. E. Stanley, {\em Fractal Concepts in Surface Growth} (Cambridge University Press, Cambridge, England, 1995).

\bibitem{canet:2010} See e.g\ L. Canet, H. Chat\'e, B. Delamotte, and N. Wschebor, Phys. Rev. Lett. {\bf 104}, 150601 (2010) and references therein.

\bibitem{nicoli:unpubl} M. Nicoli, E. Vivo, and R. Cuerno, unpublished.

\bibitem{Gallego:2007}
R. Gallego, M. Castro, and J. M. L\'{o}pez, Phys. Rev. E {\bf 76}, 051121 (2007).

\bibitem{Wio:2010_all}
H. S. Wio, J. A. Revelli, R. R. Deza, C. Escudero, and M. S. de La Lama, EPL {\bf 89}, 40008 (2010); Phys. Rev. E {\bf 81}, 066706 (2010).

\bibitem{Giada:2002}
L. Giada, A. Giacometti, and M. Rossi, Phys. Rev. E  {\bf 65}, 036134 (2002).

\bibitem{Nicoli:2008_all}
M. Nicoli, M. Castro, and R. Cuerno, Phys. Rev. E {\bf 78}, 021601 (2008); M. Nicoli, R. Cuerno, and M. Castro, Phys. Rev. Lett. {\bf 102}, 256102 (2009).

\bibitem{Canuto:1987}
C. Canuto, M. Y. Hussaini, A. Quarteroni, and T. A. Zhang, {\em Spectral Methods in Fluid Dynamics} (Springer-Verlag, New York, 1987).

\bibitem{Giacometti:2000}
A. Giacometti, and M. Rossi, Phys. Rev. E {\bf 62}, 1716 (2000).

\bibitem{siegert:1996} M. Siegert, Phys. Rev. E {\bf 53}, 3209 (1996).

\bibitem{epaps} See supplementary material at  http://link.aps.org/ supplemental/10.1103/PhysRevE.82.045202 for examples of data collapses corresponding to points on Fig.\ \ref{fig:kpz-uc}. We use $z_{\rm KPZ}=2-\alpha_{\rm KPZ} = 1.61$, as implied by Galilean invariance \cite{Barabasi:1995}.

\bibitem{chow:1995} C. C. Chow and T. Hwa, Physica D {\bf 84}, 494 (1995).

\bibitem{Boghosian:1999}
B. M. Boghosian, C. C. Chow, and T. Hwa, Phys. Rev. Lett. {\bf 83}, 5262 (1999).

\bibitem{bychkov:2000} V. V. Bychkov and M. A. Liberman, Phys. Rep. {\bf 325}, 115 (2000).

\bibitem{karlin:2004} V. Karlin, Math. Models Methods Appl. Sci. {\bf 14}, 1191 (2004).

\end{thebibliography}
\end{document}